\documentclass[pre,aps,twocolumn,showpacs,superscriptaddress,floatfix,amsmath,amssymb,nofootinbib]{revtex4}
\usepackage{amsfonts}
\usepackage{graphicx} 
\newcommand{\beq}{\begin{equation}}
\newcommand{\eeq}{\end{equation}}

\newcommand{\beqa}{\begin{eqnarray}}
\newcommand{\eeqa}{\end{eqnarray}}
\def\ra{\rangle}
\def\la{\langle}
\begin{document} 
\title{The quest for absolute zero in presence of external noise}
\author{E. Torrontegui}
\affiliation{Departamento de Qu\'{\i}mica-F\'{\i}sica, Universidad del Pa\'{\i}s Vasco - Euskal Herriko Unibertsitatea, 
Apdo. 644, Bilbao, Spain}
\affiliation{Institute of Chemistry, The Hebrew University, Jerusalem 91904, Israel}

\author{R. Kosloff}
\affiliation{Institute of Chemistry, The Hebrew University, Jerusalem 91904, Israel}
%

%

\begin {abstract}
A reciprocating quantum refrigerator is analyzed with the intention to study the limitations imposed by external noise.
In particular we focus on the behavior of the refrigerator when it approaches the absolute zero. The 
cooling cycle is based on the Otto cycle with  a working medium constituted by an ensemble of non interacting
harmonic oscillators. The compression and expansion segments are generated by
changing an external parameter in the Hamiltonian. In this case the force constant of the harmonic oscillators $m \omega^2$ is
modified from an initial to a final value. As a result, the kinetic and potential energy of the system do not commute causing 
frictional losses. By proper choice of scheduling function $\omega(t)$ frictionless solutions can be obtained in the noiseless case.
We examine the performance of a refrigerator subject to noise. By expanding from the adiabatic limit
we find that the external noise, gaussian phase and amplitude noises, reduce the amount of
heat that can be extracted  but nevertheless the zero temperature can be approached. 
\end{abstract}  	
\maketitle
\section{The system}
Reciprocating refrigerators operate by a working medium shuttling heat from the cold bath at $T_c$ temperature to the hot reservoir
with $T_h$ temperature. The task is carried out by a controlled dynamical system. By changing the Hamiltonian of the system
the internal levels of the working medium can be modified. Upon contact with the cold side the internal energy of
the working medium is forced to be lower than the equilibrium temperature $T_c$, only under these conditions heat
flows from the cold to the heat bath \cite{jahnke}. In the hot reservoir, however, a reciprocal relation is required.
In the present work
we consider a refrigerator using a controllable quantum medium as its working medium 
\cite{njp, eplY, oct, preTR, eplTR, preRT}. The medium is constituted by an ensemble of non interacting particles
bound by a shared harmonic potential whose frequency $\omega (t)$ can be controlled and varies between
two extreme values $\omega_c$ and $\omega_h$, the frequencies associated to the cold and hot
baths, respectively \cite{njp, eplY, oct}. The energy of each particle is represented by the hamiltonian
\beq
\bold{\hat H}=\frac{1}{2m}\bold{\hat P}^2+\frac{1}{2}m\omega (t)^2\bold{\hat Q}^2,
\eeq
 where $m$ is the mass of each particle, and $\bold{\hat P}$ and $\bold{\hat Q}$ are the momentum and position operators.

The dynamics of the quantum thermodynamical observables are described within the formalism of quantum open systems generated by the Liouville
superoperator, $\mathcal{L}$, studied in the Heisenberg picture \cite{lindblad, breuer},
\beq
\label{mov}
\frac{d\bold{\hat A}}{dt}=\frac{i}{\hbar}[\bold{\hat H},\bold{\hat A}]+\mathcal{L}(\bold{\hat A})+\frac{\partial \bold{\hat A}}{\partial t},
\eeq
where $\mathcal{L}$ is a generator of a completely positive Liouville superoperator.
Typically, $[\bold{\hat H}(t),
\bold{\hat H}(t')]\neq 0$ which leads to friction-like phenomena \cite{njp, preRT}, too fast adiabatic segments will generate
parasitic internal
energy which will have to be dissipated to the heat baths limiting the performance.

The considered cycle is composed \cite{njp, eplY} by two segments, termed {\it isochores} where the working medium is in contact with the cold or hot baths
and the external control field $\omega$ is maintained constant. There are also two other segments, termed {\it adiabats} where the working medium is
isolated from the baths and the frequency $\omega(t)$ is modified. This represents the quantum version of the Otto cycle. Each segment is characterized
by a quantum propagator $\mathcal{U}_s$ which maps the initial state of the working medium to the final one. At the {\it isochores} 
$\mathcal{L}=\mathcal{L}_{D}$ is the dissipative term responsible for driving the working medium to thermal equilibrium when it is in contact with either the cold or hot baths. 
For the harmonic oscillator $\mathcal{L}_{D}$ takes the explicit form \cite{njp}
\beq
\label{super}
\mathcal{L}_{D}(\hat\rho)=k_{\downarrow}(\hat a^{\dag}\hat\rho\hat a-\{\hat a\hat a^{\dag},\hat\rho\}/2)+k_{\uparrow}(\hat a\hat\rho\hat a^{\dag}-\{\hat a^{\dag}\hat a,\hat\rho\}/2),
\eeq
where $\{\hat A,\hat B\}=\hat A\hat B+\hat B\hat A$, $\hat a$ and $\hat a^{\dag}$ are the lowering and raising operators of the harmonic oscillator,
and $k_{\downarrow}$ and $k_{\uparrow}$ are the heat conductances satisfying detailed balance $k_{\uparrow}/k_{\downarrow}=\exp[-\hbar\omega/(k_{B}T)]$
being $T$ either $T_c$ or $T_h$ \cite{njp} and $k_B$ the Boltzmann's constant. At the {\it adiabats} $\mathcal{L}=\mathcal{L}_{N}$ which represents external noise in our controls.
Any realistic refrigerator is subject to fluctuations in the external control which induces noise. The main point of this paper is to
study the effects of noise on the behaviour of the Otto cycle. First we consider the phase noise, it occurs when a piecewise process
controlling the scheduling of $\omega$ in time. For such a procedure random errors are expected in the duration of the time intervals
described by the Linbland superoperator $\mathcal{L}_N$. We describe these errors by a Gaussian white noise which mathematically
is equivalent to a dephasing process on the {\it adiabats} \cite{preTR2006}. Then the dissipative operator $\mathcal{L}_N$ is given by 
\cite{gorini, breuer};
\beq
\mathcal{L}_{N_{p}}(\bold{\hat A})=-\frac{\gamma_p}{\hbar^2}[\bold{\hat H},[\bold{\hat H},\bold{\hat A}]].
\eeq
In addition to the phase noise there is a more obvious source of external noise which is induced by fluctuations in the
control frequency $\omega (t)$, this term represents Markovian random fluctuations in the profile of the harmonic oscillator. 
The dissipative Linbland term is 
\beq
\mathcal{L}_{N_{a}}(\bold{\hat A})=-\gamma_a\omega^2[\bold{\hat B},[\bold{\hat B},\bold{\hat A}]],
\eeq
where $\bold{\hat B}=m\omega\bold{\hat Q}^2/(2\hbar)$.
A more detailed description of the four segments is:

- {\it Hot isochore:} the frequency is constant $\omega=\omega_h$ and the working medium is in contact with the hot bath at temperature $T_h$. In this
case $\mathcal{L}=\mathcal{L}_D$ is the dissipative Lindblad term Eq. (\ref{super}) which leads the system toward thermal equilibrium of an
harmonic oscillator defined by $k_{\uparrow}/k_{\downarrow}=\exp[-\hbar\omega/(k_{B}T)]$ \cite{njp}. 
The dynamic of the segment are described
by the operator $\mathcal{U}_h$.

-{\it Expansion adiabat:} The working medium is isolated from the baths and the frequency $\omega (t)$ changes from the initial value $\omega_0=\omega_h$ to the
final one $\omega_f=\omega_c$. The Liouville superoperator is $\mathcal{L}=\mathcal{L}_N$ which represents external noise in our controls. The propagator
$\mathcal{U}_{hc}$ associated to this branch is our main subject of study.

- {\it Cold isochore:} the frequency is maintained constant $\omega=\omega_c$ and the working medium is in contact with the cold bath at temperature $T_c$. 
$\mathcal{L}$ is again a dissipative Lindblad term $\mathcal{L}_{D}$. The evolution operator is $\mathcal{U}_c$.

-{\it Compression adiabat:} the frequency $\omega (t)$ changes from the initial value $\omega_c$ to the
final one $\omega_h$. The Liouville superoperator $\mathcal{L}=\mathcal{L}_N$ represents the external noise. The dynamic is described by $\mathcal{U}_{ch}$.
Then the total propagator associated with one whole cycle is the product
\beq
\mathcal{U}_{cyc}=\mathcal{U}_{ch}\mathcal{U}_{c}\mathcal{U}_{hc}\mathcal{U}_{h}.
\eeq

The state that describes the system is fully characterized by the thermodynamical variables. Statistical thermodynamics define that the state is determined by the
maximum entropy condition subject to the constrains impose by the thermodynamical observables \cite{janes1, janes2, katz}. At thermal equilibrium the energy expectation
value is sufficient to represent the state of the system, maximizing the von Neumann entropy $S_{VN}=-k_BTr[\hat\rho\ln(\hat\rho)]$ subject to the energy constrain sets the 
density matrix at thermal equilibrium  \cite{katz},
\beq
\label{eq}
\hat\rho_{eq}=\frac{1}{Z}e^{-\hat{H}/(k_BT)}
\eeq
where $Z=Tr[e^{-\hat H/(k_BT)}]$ is the partition function. However, during the cycle the system is not in general at thermal equilibrium, thus in order to generalize the 
canonical form Eq. (\ref{eq}) additional thermodynamical observables are necessary to characterize the state of the system. 
The density matrix of the state that maximizes the entropy
subject to the constrains $\la\bold{\hat A}_i\ra=Tr[\bold{\hat A}_i\hat\rho]$ is
\beq
\label{Neq}
\hat\rho=\frac{1}{Z}e^{\ \sum_i\beta_i\bold{\hat A}_i}
\eeq
where $\beta_i$ are the Lagrange multipliers. This generalized canonical form is meaningful if the state can cast Eq. (\ref{Neq}) during the whole cycle leading to 
canonical invariance $\beta_j=\beta_j(t)$ \cite{katz} which is guaranteed if the set of operators $\bold{\hat A}_i$ of Eq. (\ref{Neq}) 
are closed under the dynamics generated by the equation of motion Eq. (\ref{mov}). For the part 
$i[\bold{\hat H},\bold{\hat A}]/\hbar+\partial \bold{\hat A}/\partial t$ which produces a unitary evolution it occurs that 
the hamiltonian of the system is a linear combination of the operator set $\bold{\hat A}_i$, 
$\bold{\hat{H}}(t)=\sum_m h_m\bold{\hat A}_m$ (where 
$h_m$ are the expansion coefficients) and the set $\bold{\hat A}_i$ forms a closed Lie algebra $[\bold{\hat A}_i, \bold{\hat A}_j]=\sum_k C_{ijk}\bold{\hat A}_k$ (the $C_{ijk}$ are the structure factors) 
that preserves canonical invariance \cite{levine}. For the Otto cycle the set of thermodynamical observables $\bold{\hat A}'=\{\bold{\hat P}^2, \bold{\hat Q}^2, 
\bold{\hat Q}\bold{\hat P}+\bold{\hat P}\bold{\hat Q}\}$ forms a $SU(1,1)$ closed Lie algebra. In addition, for the non-unitary part 
$\mathcal{L}$ of Eq. (\ref{mov}) this set of
three thermodynamical observables is closed to either $\mathcal{L}_{D}$, $\mathcal{L}_{N_{p}}$ and $\mathcal{L}_{N_{a}}$, which is sufficient to preserve canonical invariance \cite{njp}.
The choice of this set of thermodynamical observables is not unique, in particular we consider
the following three $\bold{\hat A}=\{\bold{\hat H}, \bold{\hat L}, 
\bold{\hat D}\}$ time dependent operators: $\bold{\hat H}=\frac{1}{2m}\bold{\hat P}^2+\frac{1}{2}m\omega (t)^2\bold{\hat Q}^2$ hamiltonian, 
$\bold{\hat L}=\frac{1}{2m}\bold{\hat P}^2-\frac{1}{2}m\omega (t)^2\bold{\hat Q}^2$ lagrangian and $\bold{\hat D}=\frac{1}{2}\omega(t)
(\bold{\hat Q}\bold{\hat P}+\bold{\hat P}\bold{\hat Q})$ momentum-space correlation which satisfy the $SU(1,1)$ algebra.
The invariant Casimir operator associated to this algebra is $\bold{\hat C}=(\bold{\hat H}^2-\bold{\hat L}^2-\bold{\hat D}^2)/(\hbar^2\omega^2)$. The closed algebra leads
to closed equations of motion on the {\it adiabats} and on the {\it isochores} \cite{eplY, preTR2003}. 

To make explicitly the connection between the state represented by the density matrix  
$\hat\rho$ and the expectation values of the three operators that guaranteed the canonical invariance let us rewrite the density matrix as \cite{njp}
\beq
\hat\rho=\frac{1}{Z}e^{\alpha\hat a^2}e^{-\beta\hat H}e^{\alpha^{*}\hat {a}^{{\dag}^2}}
\eeq
where * represents the complex conjugate. The hermiticity of the density matrix $\hat\rho=\hat\rho^{\dag}$ imposes
the real and complex nature of the time dependent Lagrange multipliers $\beta\in\mathbb{R}$ and $\alpha\in\mathbb{C}$. In terms 
of the expectation values of the thermodynamical observables $\la\bold{\hat A}_i\ra$ the Lagrange multipliers are given by \cite{njp}
\beqa
\alpha&=&\frac{2\hbar\omega[\la\bold{\hat L}\ra+i\la\bold{\hat D}\ra]}{4[\la\bold{\hat L}\ra^2+\la\bold{\hat D}\ra^2]-[\hbar\omega-2\la\bold{\hat H}\ra]^2}, \\
\beta&=&\frac{1}{\hbar\omega}\ln\bigg[\frac{4[\la\bold{\hat D}\ra^2-\la\bold{\hat H}\ra^2+\la\bold{\hat L}\ra^2]+\hbar^2\omega^2}{4[\la\bold{\hat L}\ra^2+\la\bold{\hat D}\ra^2]-[\hbar\omega-2\la\bold{\hat H}\ra]^2}\bigg].
\eeqa
\section{Equations of motion}
The dynamics generated on the {\it isochores} is obtained from Eq. (\ref{mov}):
\beq
\frac{d}{dt}
\underbrace{\left(\begin{array}{c} 
\bold{\hat H}  \\
\bold{\hat L}  \\
\bold{\hat D}
\end{array} \right)}_{=:\bold{\hat A}(t)}
=
\left(\begin{array}{rcl} 
-\Gamma &   0 & 0  \\
0  &   -\Gamma & -2\omega \\
0       &  2\omega& -\Gamma 
\end{array} \right)
\bold{\hat A}(t)+
\left(\begin{array}{c} 
\Gamma\langle\bold{\hat H}\rangle_{eq}  \\
0\\
0
\end{array} \right),
\eeq
where $\Gamma=k_{\downarrow}-k_{\uparrow}$ is the heat conductance and $\langle\bold{\hat H}\rangle_{eq}=\hbar\omega\coth[\hbar\omega/(2k_{B}T)]/2$ is the
equilibrium expectation energy. The equation of motion for the hamiltonian is decoupled from the equations for $\bold{\hat L}$
and $\bold{\hat D}$ and displays an exponential approach to  equilibrium. In contrast the lagrangian and  momentum-space 
correlation operators show an oscillatory decay to an expectation value of zero at equilibrium. More explicitly, as in the {\it isochores}
the frequency is constant, and the above equations can be  integrated,
\beqa
\bold{\hat H}(t)&=&e^{-t\Gamma}\bigg[\bold{\hat H}(0)-\bold{\hat{\mathbb{I}}}\langle\bold{\hat H_{eq}}\rangle\bigg]+\bold{\hat{\mathbb{I}}}\langle\bold{\hat H_{eq}}\rangle,
\\
\bold{\hat L}(t)&=&e^{-t\Gamma}\bigg[\cos(2\omega t)\bold{\hat L}(0)-\sin(2\omega t)\bold{\hat D}(0)\bigg],
\\
\bold{\hat D}(t)&=&e^{-t\Gamma}\bigg[\sin(2\omega t)\bold{\hat L}(0)+\cos(2\omega t)\bold{\hat D}(0)\bigg],
\eeqa
being $\bold{\hat H}(0), \bold{\hat L}(0)$ and $\bold{\hat D}(0)$ the initial values of the operators.

For the case of the {\it adiabats} the equations of motion take the form
\beq
\label{adi}
\frac{d \bold{\hat A}(t)}{\omega dt}
=
\left(\begin{array}{ccc} 
\mu+\gamma_a\omega   &   -\mu-\gamma_a\omega & 0  \\
-\mu+\gamma_a\omega  &   \mu-(4\gamma_p+\gamma_a)\omega & -2 \\
0       &  2& \mu-4\gamma_p\omega 
\end{array} \right)
\bold{\hat A}(t),
\eeq
where the three operators and $\omega$ depend on $t$ and $\mu=\dot\omega/\omega^2$ is defined as a dimensionless
adiabatic parameter \cite{eplY}. The solution of these coupled differential equations depends on the functional form of $\omega(t)$.
Now our main purpose is to solve the Eq. (\ref{adi}) for the {\it adiabats}. From this expression we write the previous $3\times 3$ matrix
$\mathcal{M}$ as $\mathcal{M}=\mathcal{M}_0+\mathcal{N}_p+\mathcal{N}_a$ where $\mathcal{M}_0$ takes into account
the noiseless evolution on the {\it adiabats} \cite{eplY} and $\mathcal{N}_p$ and $\mathcal{N}_a$ are the contributions of the phase
and amplitude noise, respectively;
\beqa
\label{M0}
\mathcal{M}_0 (t)=\left(\begin{array}{ccc} 
\mu   &   -\mu & 0  \\
-\mu  &   \mu & -2 \\
0       &  2& \mu
\end{array} \right),
\eeqa
\beqa
\nonumber
\mathcal{N}_p (t)=-4\gamma_p\omega
\left(\begin{array}{ccc} 
0   &   0& 0  \\
0 &   1& 0 \\
0       &  0 & 1
\end{array} \right),
\quad
\mathcal{N}_a (t)=\gamma_a\omega
\left(\begin{array}{ccc} 
1   &   -1 & 0  \\
1  &   -1 & 0 \\
0       &  0& 0
\end{array} \right).
\eeqa
The solution for the noiseless evolution has been obtained before \cite{eplY}. Usually the time evolution with an arbitrary functional
form of $\omega (t)$ will involve  quantum friction \cite{njp} due to the resultant parasitic increase in the internal energy. The dissipation of this energy, in particular into the cold bath, can cause the refrigerator to stop cooling.
But it is known that there are some special functional forms of $\omega (t)$ that allow a frictionless evolution in the
{\it adiabats} \cite{eplY, oct, prlXi, 3D, adolfoOTTO}. In particular we consider the frictionless $\mu=const$ \cite{eplY} case which sets a frequency profile 
\beq
\label{omega}
\omega (t)=\omega_0/
(1-\mu\omega_0 t),
\eeq
being $\omega_0$ the initial frequency at the {\it adiabats}.
To get the noiseless solution for  constant $\mu$ we seek a solution of the type $\mathcal{U}_{hc}^{0}=\mathcal{U}_1\mathcal{U}_2$ where $[\mathcal{U}_1,\mathcal{U}_2]=0$ and being $\mathcal{U}_1$ the part associated with the adiabatic
evolution. The operator $\mathcal{U}_1$ is obtained by changing the time variable $d\theta=\omega (t)dt$ and
 factoring out the term $\mu{\bold{\hat{\mathbb{I}}}}$ in 
$\mathcal{M}_0$ thus 
\beq
\label{U1}
\mathcal{U}_1=\frac{\omega (t)}{\omega_0}{\bold{\hat{\mathbb{I}}}}, \quad \omega_0=\omega (t=0).
\eeq
The non-adiabatic contribution is codified into the $\mathcal{U}_2$ operator which is obtained by diagonalizing the 
extra-diagonal terms for $\mu < -2$
\beqa
\label{U2}
\mathcal{U}_2=\frac{1}{\Omega^2}\left(\begin{array}{ccc} 
4-c\mu^2   &   -\mu\Omega s& -2\mu(c-1)  \\
-\mu\Omega s &   \Omega^2 c & -2\Omega  s\\
2\mu(c-1)       &  2\Omega s & 4c-\mu^2
\end{array} \right),
\eeqa
where $\Omega=\sqrt{4-\mu^2}$, $c=\cos(\Omega\theta)$, $s=\sin(\Omega\theta)$ and $\theta (t)=-\ln(\frac{\omega_0}{\omega(t)})/
\mu$. 
From Eq. (\ref{U2}) we see that the propagator $\mathcal{U}_2$ induces mixing of $\bold{\hat H}$, $\bold{\hat L}$ and $\bold{\hat D}$.
To characterize the deviation from perfect factorization of the three operators we define an adiabaticity measure $\delta$
\cite{preRT} as
\beq
\label{delta}
\delta=[\mathcal{U}_{1}^{-1}\mathcal{U}_{hc}](1,1)-1,
\eeq
in this context of noiseless dynamics and constant $\mu$,  $\delta=\mathcal{U}_{2}(1,1)-1$.
The propagator $\mathcal{U}_2$ is to be expanded around $\mu=0$ which corresponds to the adiabatic limit.
For $\mu < -2$, $\mathcal{U}_2$ corresponds to a rotation matrix so $\bold{\hat H}$, $\bold{\hat L}$ and $\bold{\hat D}$ describes a periodic
motion. Each period is defined by 
\beq
X=\Omega\theta=2n\pi, \quad n=0,1, 2 ,3 \dotsc,
\eeq
thus at the end of each one $\mathcal{U}_2$ restores the identity matrix and the perfect adiabatic following conditions $\delta=0$
are satisfied, leading to frictionless dynamic. 
This frictionless condition sets a quantization on the constant adiabatic parameter $\mu$,
\beq
\label{muq}
\mu=\frac{-2\ln\left(\frac{\omega_0}{\omega_f}\right)}{\sqrt{4n^2\pi^2+\ln^2\left(\frac{\omega_0}{\omega_f}\right)}},
\eeq
where $\omega_f$ is the final frequency at the {\it adiabats}.
The frictionless allocating times in the {\it adiabats} can be calculated from Eqs. (\ref{omega}) and (\ref{muq}),
\beq
\label{tau}
\tau=\frac{\left(\frac{\omega_0}{\omega_f}-1\right)\sqrt{4n^2\pi^2+\ln^2\left(\frac{\omega_0}{\omega_f}\right)}}{2\omega_0
\ln\left(\frac{\omega_0}{\omega_f}\right)}.
\eeq
\section{Effect of noise}

The propagators for both phase and amplitude noise are drived. For a constant $\mu$ the exact solutions
can be obtained by numerical integration of Eq. (\ref{adi}). To gain insight
the exact solution is compared to approximations which provide more transparent physical interpretations. 
To this end we consider separately the phase 
and amplitude noise in the adiabatic frictionless $\mu\rightarrow 0$ limit.
\subsection{Phase noise}

Now we include the phase noise trough the dissipative operator $\mathcal{L}_{N_{p}}(\bold{\hat A})=-\frac{\gamma_p}{\hbar^2}[\bold{\hat H},[\bold{\hat H},\bold{\hat A}]]$ and we seek a solution of the product form $\mathcal{U}_{hc}=\mathcal{U}_{1}\mathcal{U}_{2}
\mathcal{U}_{3}$. The equations of motion for $\mathcal{U}_3$ are obtained from the interaction picture
\beqa
\nonumber
\frac{d}{\omega dt}\mathcal{U}_{3p}(t)&=&\mathcal{U}_2(-t)
\mathcal{N}_p(t)
\mathcal{U}_2(t)\mathcal{U}_{3p}(t)
\\
&=&\mathcal{W}_p(t)\mathcal{U}_{3p}(t), \label{difWp}
\eeqa
where
\begin{widetext}
\beq
\label{Wpexact}
\mathcal{W}_p(t)=\frac{4\gamma_p\omega(t)}{\Omega^4}
\left(\begin{array}{ccc} 
(c-1)\mu^2[\mu^2(1+c)-8] &\mu s(\mu^2 c-4)\Omega &2\mu (c-1) (\mu^2 c-4)\\
-\mu s(\mu^2 c-4)\Omega  &   -[\mu^2(s^2-1)+4]\Omega^2 & -2\mu^2 s(c-1)\Omega \\
-2\mu (c-1) (\mu^2 c-4) & -2\mu^2 s(c-1)\Omega & -[\mu^2(\mu^2-4s^2-8c)+16]
\end{array} \right). 
\eeq
\end{widetext}
The Magnus expansion \cite{blanes} is employed to solve Eq. (\ref{difWp}) which can be rewritten as $\frac{d\mathcal{U}_{3p}(X)}{dX}=\mathcal{W}_p(X)\mathcal{U}_{3p}(X)/\Omega$ and  obtain the $n$ period propagator $\mathcal{U}_{3p}(X=2n\pi)$, 
\beq
\mathcal{U}_{3p}(X=2n\pi)\approx e^{B_1+B_2+\dotsc}
\eeq
where $B_1=\int_{0}^{2n\pi} dX\mathcal{W}_{a}(X)/\Omega$, $B_2=\frac{1}{2}\int_{0}^{2n\pi}\int_{0}^{X} dX dX' [\mathcal{W}_{a}(X),\mathcal{W}_{a}(X')]/\Omega^2$ and so on. The first order Magnus term leads to
\beqa
B_{1p}&=&\frac{\gamma_p\omega_h\mu\big(-1+e^{2n\pi\mu/\Omega}\big)}{-16+3\mu^2} \nonumber\\
&\times&
\left(\begin{array}{ccc} 
-32  &  -16 & -\frac{32}{\mu}-6\mu\\
16   &   -4+\frac{64}{\mu^2}  & 6\mu \\
\frac{32}{\mu}+6\mu& 6\mu &   12+\frac{64}{\mu^2}
\end{array} \right).
\eeqa
In the adiabatic limit we neglect the $\mathcal{O}(\mu)$ terms in the non-diagonal and the  $\mathcal{O}(0)$ terms in the diagonal so
\beq
\mathcal{U}_{3p}(X=2n\pi)=
\left(\begin{array}{ccc} 
e^{\gamma_p\mathcal{F}\mu}  &  0 & 0\\
0  &  \frac{4}{4+\mu^2}e^{-4\gamma_p\mathcal{F}/\mu} &0 \\
0  &   0  & e^{-\gamma_p\mathcal{F}\Omega^2/\mu}
\end{array} \right),
\eeq
where $\mathcal{F}=-\frac{16\omega_h}{(-16+3\mu^2)}\big(-1+e^{2n\pi\mu/\Omega}\big)$,
then the frictionless parameter given by Eq. (\ref{delta}) for phase noise is
$\delta=U_3(1,1)-1$,
\beq
\label{dp}
\delta_p\approx -1+e^{\gamma_p\mathcal{F}\mu}.
\eeq
notice that as $\mu \rightarrow 0 $  the phase noise in the first order Magus approximation vanishes $\delta_p \rightarrow 0$.

When computing the second order Magnus expansion for phase noise we get 
for the adiabatic limit 
\beq
\label{U3p2}
\mathcal{U}_{3p}^{(2)}=
\left(\begin{array}{ccc} 
\cosh \beta & -\sinh \beta   & 0\\
-\sinh \beta      &   \cosh \beta     & 0 \\
0 &    0 & 1
\end{array} \right),
\eeq
where $\beta= \frac{16\omega_h^2\gamma_p^2}{(4+3\mu^2)}\big(-1+e^{4n\pi\mu/\Omega}\big)$, so $\delta_p^{(2)}=-1+\cosh\beta$. When $n\rightarrow\infty$ it sets a finite value for the influence of phase noise
\beq
\delta_p^{(2)}=-1+\cosh[4\gamma_p^2(\omega_h^2-\omega_c^2)].
\label{p22}
\eeq

\subsection{Amplitude noise}
We progress in a similar way to calculate the phase noise produced by the dissipative operator 
$\mathcal{L}_{N_{a}}(\bold{\hat A})=-\gamma_a\omega^2[\bold{\hat B},[\bold{\hat B},\bold{\hat A}]]$. In this case the
interaction picture for $\mathcal{U}_{3a}$ is
\beqa
\nonumber
\frac{d}{\omega dt}\mathcal{U}_{3a}(t)&=&\mathcal{U}_2(-t)
\mathcal{N}_a(t)
\mathcal{U}_2(t)\mathcal{U}_{3a}(t)
\\
&=&\mathcal{W}_a(t)\mathcal{U}_{3a}(t), \label{difWa}
\eeqa
being
\begin{widetext}
\beq
\label{Waexact}
\mathcal{W}_a(t)=\frac{\gamma_a\omega(t)}{\Omega^4}
\left(\begin{array}{ccc} 
(4-\mu^2c+\mu s\Omega)^2 & (\mu s+\Omega c)(-4+\mu^2 c-\mu\Omega s)\Omega  & 2(\mu-\mu c+\Omega s)(4-\mu^2 c+\mu\Omega s) \\
(\mu s+\Omega c)(4-\mu^2 c+\mu\Omega s)\Omega   &   -(\mu s +\Omega c)^2\Omega^2 & 2(\mu s+\Omega c)(\mu-\mu c+\Omega s)\Omega \\
-2(\mu-\mu c+\Omega s)(4-\mu^2 c+\mu\Omega s)  &    2(\mu s+\Omega c)(\mu-\mu c+\Omega s)\Omega  & -4(\mu-\mu c+\Omega s)^2
\end{array} \right).
\eeq
\end{widetext}
In this case the first order Magnus term leads to
\beqa
B_{1a}&=&\frac{\gamma_a\omega_h\big(-1+e^{2n\pi\mu/\Omega}\big)}{-16+3\mu^2} \nonumber\\
&\times&
\left(\begin{array}{ccc} 
-\frac{16}{\mu}+\mu  &  -\mu &4 \\
\mu   &   \frac{8}{\mu}-\mu & 2 \\
4& 2 &   \frac{8}{\mu}
\end{array} \right),
\eeqa
and in the adiabatic limit just the $\mathcal{O}(\mu^{-1})$ terms are considered, then
\beq
\mathcal{U}_{3a}(X=2n\pi)=
\left(\begin{array}{ccc} 
e^{\gamma_a\mathcal{F}/\mu}  &  0 & 0\\
0  &  e^{-\gamma_a\mathcal{F}/(2\mu)} &0 \\
0  &   0  & e^{-\gamma_a\mathcal{F}/(2\mu)}
\end{array} \right),
\eeq
so the adiabaticity measure reads
\beq
\label{da}
\delta_a\approx -1+e^{\gamma_a\mathcal{F}/\mu}.
\eeq
%

\begin{figure}[t]
\begin{center}
\includegraphics[width=8.cm]{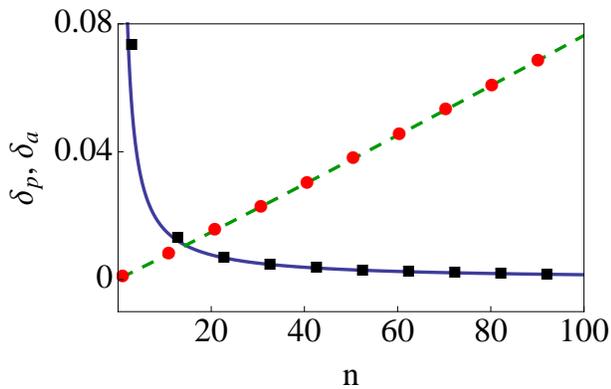}
\caption{(Color online) Parameter $\delta$ versus the frictionless cycle number $n$ for phase and amplitude noise. (Blue solid line) approximated $\delta_p$ computed
with the analytic Eq. (\ref{dp}), (black squares) exact $\delta_p$ solving numerically Eqs. (\ref{difWp}) and (\ref{Wpexact}).
(Green dashed line) approximated $\delta_a$ computed
with the analytic Eq. (\ref{da}), (red circles) exact $\delta_a$ solving numerically Eqs. (\ref{difWa}) and (\ref{Waexact}).
Parameter values: $\omega_c=2\pi\times1000$ rad/s, $\omega_h/\omega_c=25$, $\gamma_p=10^{-6}s^{-1}$ and 
$\gamma_a=5\cdot10^{-9}s^{-1}$.
}
\label{f1}
\end{center}
\end{figure}
%
%
%
%
%
%
These two analytic expressions for $\delta$ match the exact values obtained numerically solving Eqs. (\ref{difWp}) or (\ref{difWa}) for phase and amplitude noise as we infer from Fig. \ref{f1}. Amplitude and phase noise show the opposite behavior
in the adiabatic limit of  $n\rightarrow\infty$. Amplitude noise diverges when the process is more adiabatic $\delta_a \sim 1/\mu$
Eq. (\ref{da}),  whereas phase noise $\delta_p$ tends to a constant as $\mu \rightarrow 0$ Eq. (\ref{p22}). 
\section{Minimal temperatue}
Does noise in the Otto cycle have any consequence on the performance of the refrigerator
when $T_c\rightarrow 0$? Is there a minimum temperature above the absolute zero
where the refrigerator stops working? 

The  conditions for refrigeration require that on the cold side the internal energy of the working medium is
smaller than the equilibrium energy with the cold bath at the end of the {\it expansion adiabat},
\beq
\langle\bold{\hat H}\rangle_c\leq\langle\bold{\hat H}\rangle_{eq}(T_c)=\frac{\hbar\omega_c}{2}\coth\left(\frac{\hbar\omega_c}{2k_BT_c}\right).
\eeq
On the {\it hot isochore} the lowest temperature that can be obtained is in equilibrium 
\beq
\langle\bold{\hat H}\rangle_h\geq\langle\bold{\hat H}\rangle_{eq}(T_h)=\frac{\hbar\omega_h}{2}\coth\left(\frac{\hbar\omega_h}{2k_BT_h}\right).
\eeq
Under these conditions of thermal equilibrium $L=D=0$ \cite{njp} and independently of what kind of noise we consider the change in 
$\langle\bold{\hat H}\rangle$ along the {\it expansion adiabat} is
\beq
\label{evolH}
\frac{\langle\bold{\hat H}\rangle_c}{\omega_c}\approx (1+\delta)\frac{\langle\bold{\hat H}\rangle_h}{\omega_h},
\eeq
being $\delta$ the deviation from the perfect adiabatic following defined in Eq. (\ref{delta}). Then the maximum
heat that can be extracted at the {\it adiabats} is given by
\beqa
\nonumber
\mathcal{Q}(max)&=&\langle\bold{\hat H}\rangle_{eq}(T_c)-\langle\bold{\hat H}\rangle_c
\\
&\approx&\mathcal{Q}_0(max)-\frac{\hbar\omega_c}{2}\delta\coth\left(\frac{\hbar\omega_h}{2k_BT_h}\right), \label{Qmax}
\eeqa
where $\mathcal{Q}_0(max)=\frac{1}{2}\hbar\omega_c\left[\coth\left(\frac{\hbar\omega_c}{2k_BT_c}\right)-\coth\left(\frac{\hbar\omega_h}{2k_BT_h}\right)\right]$, is the maximum heat that can be extracted without noise \cite{njp}.
Noise decreases the heat
extraction efficiency. 
The condition for refrigeration is that $\mathcal{Q}(max)\geq 0$, then the temperature should satisfy
\beq
\label{Tc}
Tc\geq\frac{\hbar\omega_c}{2k_B arccoth\big[(1+\delta)\coth\big(\frac{\hbar\omega_h}{2k_BT_h}\big)\big]},
\eeq
where $\delta=\mathcal{U}_{3}(1,1)-1$. For a noiseless evolution $\delta=0$, it sets the Carnot limit
\beq 
\label{Tcarnot}
T_c\geq\frac{\omega_c}{\omega_h}T_h.
\eeq
In contrast if phase or amplitude noise is present during the evolution the frictionless parameter $\delta$ is different from 0, 
the value of $\delta$ in Eq. (\ref{Tc}) should be replaced by Eq. (\ref{dp}) or (\ref{da}) depending what kind of noise is considered.
%
%
%
\begin{figure}[t]
\begin{center}
\includegraphics[width=7.3cm]{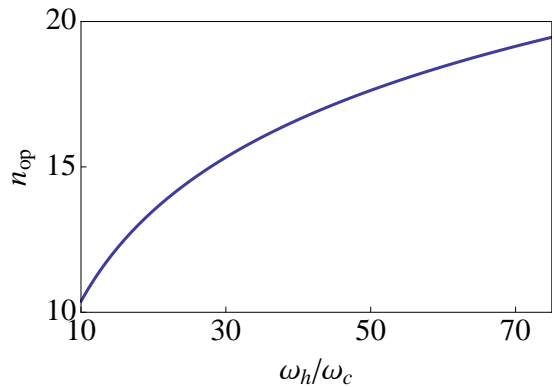}
\caption{(Color online) Optimum cycle $n_{op}$ versus the ratio $\omega_h/\omega_c$. Same parameter values as in Fig. \ref{f1}.
}
\label{f2}
\end{center}
\end{figure}

When both amplitude and phase noise are present there is an optimum cycle index $n$ which minimizes the temperature due to the two opposite
tendencies when the process becomes more adiabatic, see Fig. \ref{f1}.  In that case the evolution operator $\mathcal{U}_3$ satisfies
\beqa
\nonumber
\frac{d}{\omega dt}\mathcal{U}_{3pa}(t)&=&\mathcal{U}_2(-t)
[\mathcal{N}_p(t)+\mathcal{N}_a(t)]
\mathcal{U}_2(t)\mathcal{U}_{3pa}(t)
\\
&=&\mathcal{W}_{pa}(t)\mathcal{U}_{3pa}(t),
\eeqa
where $\mathcal{W}_{pa}(t)=\mathcal{W}_{p}(t)+\mathcal{W}_{a}(t)$. 
Since the commutator $[U_{3p},U_{3a}]=0$ then
$U_{3pa}= U_{3p}U_{3a}$ and as $U_{3p}$ and $U_{3a}$ are diagonals $U_{3pa}(1,1)= U_{3p}(1,1)U_{3a}(1,1)$ or equivalently using the definition
of Eq. (\ref{delta}), $U_{3pa}(1,1)=(\delta_p+1)(\delta_a+1)$ so
\beq
\delta_{pa}\approx\delta_p+\delta_a.
\eeq
To find the optimum $n_{op}$ cycle we solve $\partial\delta_{pa}/\partial n=0$, the result is plotted in Fig. \ref{f2}.
This point $n_{op}$ correspond to the intersection $\delta_p(n_{op})=\delta_a(n_{op})$. Equating Eqs. (\ref{dp}) and (\ref{da}) we find that
$\mu(n_{op})=\sqrt{\gamma_a/\gamma_p}$ or explicitly using Eq. (\ref{muq}),
\beq
\label{nop}
n_{op}=\frac{1}{2\pi}\sqrt{\frac{4\gamma_p}{\gamma_a}-1}\ln\bigg(\frac{\omega_h}{\omega_c}\bigg).
\eeq
Replacing Eq. (\ref{nop}) into Eq. (\ref{dp}) or (\ref{da}) we obtain the optimal $\delta$ value
\beq
\delta_{pa}(n_{op})=-1+e^{-\frac{16\sqrt{\gamma_a}\gamma_p^{3/2}(\omega_h-\omega_c)}{3\gamma_a-16\gamma_p}}.
\eeq
Inserting this expression into Eq. (\ref{Tc}) we get the minimum temperature
\beq
Tc\geq\frac{\hbar\omega_c}{2k_B arccoth\big[e^{-\frac{16\sqrt{\gamma_a}\gamma_p^{3/2}(wh-wc)}{3\gamma_a-16\gamma_p}}\coth\big(\frac{\hbar\omega_h}{2k_BT_h}\big)\big]},
\eeq
we see that if $\omega_c\rightarrow 0$ then $T_c\rightarrow 0$, thus
in contrast with  reciprocating refrigerators which use different working mediums possessing a gap in their spectrum
leading to a critical temperature in 
which they stop working due to the noise effects \cite{eplTR, preTR}. In this model  by decreasing the gap $\omega_c \rightarrow 0$ the
zero temperature $T_c=0$ is approachable  even when the phase and amplitude noises are taken into account.
\section{Conclusions}
We have analyzed the influence of external noise on the behavior of the  Otto cycle constituted by a quantum harmonic heat engine when it approaches the absolute zero. Phase and
amplitude noise reduces the amount of heat that can be extracted per cycle showing a different behavior in the
adiabatic limit, phase noise tends to a constant value whereas amplitude noise diverges. This two opposite tendencies 
set an optimal cycle for which the absolute zero can be approached decreasing the energy gap $\omega_c\rightarrow 0$ when both kind of noises are considered together. 

The engine performance is examined for the particular frictionless trajectories $\mu=const$, however these trajectories are not unique \cite{prlXi, transport, 3D, adolfoOTTO} and in fact their are infinite possibilities for scheduling
$\omega(t)$. Given this freedom for choosing a frictionless trajectory, optimal control theory can be applied \cite{eplY, oct, stef,BECtransport, XiOCT}. For noiseless evolution \cite{eplY, oct} bang-bang
trajectories are the optimal to produce the cycle in the smallest time, how does noise modify these trajectories?

The dynamics of our reciprocating quantum refrigerator is codified into Eq. (\ref{M0}). Note that $\mathcal{M}_0 (t)$ has an exceptional point \cite{kato,raam} at $\mu=\pm 2$ where all the eigenvalues and eigenvectors
of the matrix collapse to the same one. A better physical understanding of these points and their relation with the speed limits are open questions.
\acknowledgements{  
We are grateful to Y. Rezek 
and  A. Levy for fruitful discussions.  
We acknowledge funding by Projects No.  IT472-10, No. FIS2009-12773-C02-01, No. FIS2012-36673-C03-01,
UPV/EHU program UFI 11/55 and the Israel Science Foundation (ISF).}


\begin{thebibliography}{10}
\bibitem{jahnke} T. Jahnke, J. Birjukov and G. Mahler, Ann. Phys. (Leipzig), \textbf{17,} (2008) 88.
\bibitem{njp} Y. Rezek and R. Kosloff, New Journal of Physics \textbf{8,} 83 (2006).
\bibitem{eplY} Y. Rezek, P. Salomon, K. H. Hoffman and R. Kosloff, Euro. Phys. Lett. \textbf{85,} (2009) 30008.
\bibitem{oct} P. Salomon, K. H. Hoffman, Y. Rezek and R. Kosloff, Phys. Chem. Chem. Phys. \textbf{11,} 1027 (2009).
\bibitem{preTR} T. Feldmann and R. Kosloff, Phys. Rev. E \textbf{70,} (2004) 046110.
\bibitem{eplTR} T. Feldmann and R. Kosloff, Euro. Phys. Lett \textbf{89,} (2010) 20004.
\bibitem{preRT} R. Kosloff and T. Feldmann, Phys. Rev. E \textbf{82,} 011134 (2010).
\bibitem{lindblad} G. Lindblad, Commun. Math. Phys. \textbf{48,} 119 (1976).
\bibitem{breuer} H. P. Breuer and F. Ptruccione, {\it Open Quantum Systems} (Oxford University Press, New York, 2002).
\bibitem{preTR2006} T. Feldmann and R. Kosloff, Phys. Rev. E \textbf{73,} 025107(R) (2006). 
\bibitem{gorini} V. Gorini and A. Kossakowski, J. Math. Phys. \textbf{17,} 1298 (1976)
\bibitem{janes1} E. T. Janes, Phys. Rev. {\bf 106} 620 (1957).
 \bibitem{janes2} E. T. Janes, Phys. Rev. {\bf 108} 171 (1957).
\bibitem{katz} A. Katz {\it Principles of Statistical Mechanics. The Information Theoretic Approach} (1967).
\bibitem{levine} Y. Alhassid and R. D. Levine, Phys. Rev. A \textbf{18,} 189 (1978). 
\bibitem{preTR2003} T. Feldmann and R. Kosloff, Phys. Rev. E \textbf{68,} 016101 (2003).
\bibitem{prlXi} X. Chen, A. Ruschhaupt, S. Schmidt, A. del Campo, D. Gu\'ery-Odelin, and J. G. Muga, Phys. Rev. Lett. 
\textbf{104,} 063002 (2010).
\bibitem{3D} E. Torrontegui, 
X. Chen, M. Modugno, A. Ruschhaupt, D. Gu\'ery-Odelin, and J. G. Muga, 
Phys. Rev. A \textbf{85}, 033605 (2012).
\bibitem{adolfoOTTO} A. del Campo, J. Goold, and M. Paternostro, arXiv:1305.3223 (2013).
\bibitem{blanes} S. Blanes, F. Casas, J. A. Oteo and J. Ros, Phys. Rep. \textbf{470,} 151 (2009).
\bibitem{transport} E. Torrontegui, S. Ib\'a\~nez, Xi Chen, A. Ruschhaupt, D. Gu\'ery-Odelin, and J. G. Muga,
Phys. Rev. A \textbf{83}, 013415 (2011).
\bibitem{stef} D. Stefanatos, J. Ruths, and J.-S. Li, Phys. Rev. A \textbf{82}, 063422 (2010).
\bibitem{BECtransport} E. Torrontegui, X. Chen, M. Modugno, S. Schmidt, A. Ruschhaupt, and J. G. Muga,
New Journal of Phys \textbf{14,} 013031 (2012).
\bibitem{XiOCT} X. Chen, E. Torrontegui, D. Stefanatos, J. S. Li and J. G. Muga, Phys. Rev. A \textbf{84,} 043415 (2011).
\bibitem{kato} T. Kato, {\it Perturbation theory of linear operators} (Springer, Berlin, 1966). 
\bibitem{raam} R. Uzdin, E. Dalla Torre, R. Kosloff, N. Moiseyev,  arXiv:1212.3077 (2012).
\end{thebibliography}
\end{document}